# Towards studying Service Function Chain Migration Patterns in 5G Networks and beyond

Rami Akrem Addad[1], Diego Leonel Cadette Dutra[2], Miloud Bagaa[1], Tarik Taleb[1,4] and
Hannu Flinck[3]

[1] Aalto University, Espoo, Finland

[2] Federal University of Rio de Janeiro, Rio de Janeiro, Brazil

[3] Nokia Bell Labs, Espoo, Finland

[4] Centre for Wireless Communications (CWC), University of Oulu, Oulu, Finland.

*Abstract*—Given the indispensable need for a reliable network architecture to cope with 5G networks, 3GPP introduced a covet technology dubbed 5G Service Based Architecture (5G-SBA). Meanwhile, Multi-access Edge Computing (MEC) combined with SBA conveys a better experience to end-users by bringing application hosting from centralized data centers down to the network edge, closer to consumers and the data generated by applications. Both the 3GPP and the ETSI proposals offered numerous benefits, particularly the ability to deliver highly customizable services. Nevertheless, compared to large datacenters that tolerate the hosting of standard virtualization technologies (Virtual Machines (VMs) and servers), MEC nodes are characterized by lower computational resources, thus the debut of lightweight micro-service based applications. Motivated by the deficiency of current micro-services-based applications to support users' mobility and assuming that all these issues are under the umbrella of Service Function Chain (SFC) migrations, we aim to introduce, explain and evaluate diverse SFC migration patterns. The obtained results demonstrate that there is no clear vanquisher, but selecting the right SFC migration pattern depends on users' motion, applications' requirements, and MEC nodes' resources.

## I. Introduction

The 3GPP has adopted a new architecture, based on microservices and web principles, dubbed 5G-SBA [1]. The SBA allows the 5G network to be flexible, reusable, and customizable, as it leverages on network functions (NFs) [2]. Having such a strong proposal derives the necessity of an efficient orchestration system where Network Function Virtualization (NFV) and Software-Defined Networking (SDN) are expected to be a key future target for allowing a fast and reliable NFs' programmability [3]. Nonetheless, among new industry use cases targeted by the 5G, there exist scenarios that go beyond what the current device-centric mobility approaches can support. The mobility of low latency communication services, shared by a group of moving devices, e.g., autonomous vehicles that share sensor data, is a prime example of these cases. These use-cases' demands for ultra-low latency can be addressed by leveraging the MEC concept [4]. By allowing the instantiation of applications nearby to the network edge, in the vicinity of users, MEC is acknowledged as one of the key pillars for meeting the demanding Key Performance Indicators (KPIs) of 5G [5].

However, compared to large data-centers that tolerate the hosting of standard virtualization technologies (VMs and servers), MEC nodes are characterized by lower computational resources. Furthermore, different standards development organizations are heavily pushing towards adopting microservices approaches and architectures [6], [7]. Therefore, when compared to traditional VMs [8] based on quick deployment, startup time, fast replication, live service migration, and scaling methods, container technologies form the ideal alternative for both MEC environments and emerging concepts of micro-services.

Both 3GPP and ETSI proposals offered numerous benefits, particularly the reduction of the network latency. However, users nowadays are everything except motionless, which induces a serious lack of flexibility and may take users far away from the original MEC node where their service started running, to overcome this problem, a new concept, dubbed Follow Me Cloud (FMC) [9], [10], has been introduced. The FMC permits services' movabilities amid different MEC nodes while ensuring low latency communications to end-users, as an FMC is a single instance moving in concordance with the end-user. Moreover, the type of services running in the Data Network (DN), which was ignored by telecommunication standardization entities, is expected to be a micro-service one. Therefore, as modern services may expand over multiple MECs, which introduces new issues – the management of instances on different MECs instead of one compared to the FMC – to ensure service continuity, links between the instances forming distributed MEC applications, additionally to links related to end-users, must be taken into account. Based on these observations, and assuming that all these issues are under SFC's migration umbrella, the contributions of this paper can be summarized as follows:

- The introduction of four practical SFC migration patterns to support micro-service based applications in the DN part from the proposed combined architecture of 3GPP and ETSI;
- A detailed evaluation of the proposed patterns, where different criteria will be considered to validate the new suggested type of migrations;

- A final comparison is presented to determine the most suitable SFC migration pattern within the 5G network.

The remainder of this paper is organized as follows. Section II outlines the related works. Various SFC migrations patterns with their respective design overview and the suitable implementation are presented in Section III. Section IV illustrates the experimental setup and discusses the obtained results. Finally, we conclude the paper and introduce future research challenges in Section V.

## II. RELATED WORK

Machen et al. [11] presented a multi-layer framework for migrating active applications in the MEC, their results show reduced total migration times, the downtime was considerable with an average of 2s in case of a blank container. The increase of the downtime is due to the non-use of the iterative approach in the live migration process. The authors of [12] proposed and evaluated three different mechanisms to improve the enduser experience by using the container-based live migration, their results show the efficiency of these solutions compared to prior works. Addad et al. [13] presented a framework for managing reliable live migrations of virtual resources across different Infrastructure as a Service (IaaS), handling unexpected cases while ensuring high QoS and a very low downtime without human intervention. The authors considered the inter-cloud migration by leveraging the SDN technology for traffic steering and re-direction, in addition to multiple migration processes.

Sun et al. [14] investigated how to migrate multiple correlated instances of VMs, defining the relationship only among concurrent migrated VMs. Haikun and Bingsheng [15] have presented a Coordinated Live Migration of Multi-Tier Applications in Cloud Environments, they detailed the difference of a single-VM migration, compared to VMs in a multi-tier application followed by a formulation of a correlated VM migrations problem. The authors designed and implemented a coordination system that can be used as a basis for enabling one of the desired strategies of SFC migration related to the network control part. However, the authors did not investigate the use of micro-services based technologies (containers) that are expected to be playing an essential role in the 5G and beyond networks.

With respect to the previously cited works, in this study, we introduce complete SFC migrations patterns, the SDN implication, and the inter-cloud live migration. Seeing that new use-cases entrance will beget a highly mobile environment and reduce the latency, this work is a must for achieving the 1 ms latency objective for the upcoming 5G mobile systems and beyond.

## III. SERVICE FUNCTION CHAINS MIGRATIONS PATTERNS FOR BEYOND 5G NETWORKS

### A. Main architecture and problem formulation

Usually, a three-layer cloud-based architecture can be represented as a general 5G architecture, where the core layer is a robust computing power setup from different vendors, e.g., Azure, Rackspace, and private clouds based on OpenStack, while the MEC layer hosts container-based technologies, e.g., LXC, LXD, Docker, and Rocket, given the insufficiency of computational resources to serve the users layer. For the simplicity's sake, we omit the core layer in this representation, moreover, we can host the MEC orchestrator in the core layer to allow a global view of all entities present in the MEC layer. Normally, the locations of the DNs and the User Plane Functions (UPFs) are the choice of the network operator. Though, because of a lack of trust between operators and to guarantee the most common architecture in a real case, the first deployment scenario presented in [2] is adopted. We assume that we have a connected car management scenario, i.e., it can be a drone-based management scenario as well, the connected car moves from a location to another one, from MEC1 to MEC2 in Fig. 1. Initially, the car is served by a set of network functions (NFs) in perfect coordination and synchronization that form an SFC, i.e., Service Function Chain 1, in Fig. 1. This SFC1 can deliver a secure video streaming service whither the $n-1$ NFs are security checks as firewall and IPS; while the remaining NF is the video server streamer. To follow end-users' mobility, the SFC needs to be shifted away, i.e., live migrated, while conserving all links and communication between NFs forming the moving SFC. The main focus is to implement the SFC migration patterns, to ensure a seamless migration across MEC nodes, without taking into account other use-case-specific aspects, such as the signal strength received by each vehicle, user equipment (UE) or UAV, and the traffic steering done by the UPF within the 3GPP domain.

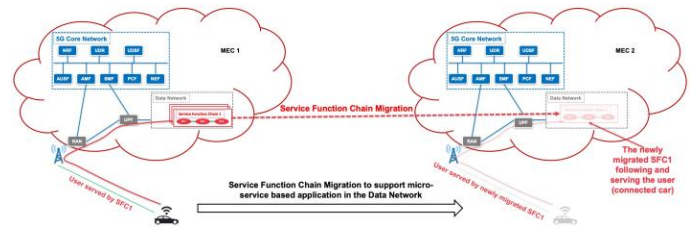

Fig. 1: Service Function Chain Migration to support microservice based application in the Data Network

To validate our proposed architecture, we need to synchronize multiple live migrations. Initially, we start our blueprint by presenting all the envisaged SFC migration scenarios, how we synchronize NFs' migration, what is the gain and the different constraints and finally decide the approach to ratify to meet 5G's low latency requirements based on evaluation in Section IV.

*B. Asynchronous State-full Service Function Chain Migration*

In this type of SFC migration, we start unsupervised live migrations for each SFC's NF, and as the last live migration end, we finish the SFC migration. Then, we can reestablish the NFs' network connectivity. We use this scenario as a worstcase upper-bound to evaluate the computational, i.e., CPU, RAM, and DISK and communication network resources, i.e., delay, and bandwidth consumption for the SFC migration.

*C. Synchronized State-full Service Function Chain Migration*

The first approach is considered a trivial solution that may consume all types of available resources, thus, we introduce the synchronized SFC migration. The well known live migration process usually takes four steps, disk copy, non-blocking memory copy (pre-dump actions in CRIU [16]), final blocking memory copy (dump action in CRIU), and restore. While we can do the first two steps without stopping the virtualized instances, the third step must freeze containers until the final step restore it afterward. Thus, a synchronized SFC migration approach aims to efficiently control each step separately, as this fine-grained control reduces the overall system resource consumption. Albeit different strategies can be employed to eliminate the system overhead caused by multiple coordinated and parallel migration processes, we selected two patterns to be presented, for both patterns we consider an SFC with two NFs:

*1) Synchronized Wait-For-Me Pattern:* In this strategy, we allow the first and second steps of the migration process to run in parallel, and we have a barrier just before the final memory blocking action, i.e., dump. Then both instances have to wait to continue their migration process. We can observe the benefits of this approach in scenarios with plenty of network resources. However, as the size of the virtualizations instances is rarely the same or even equivalent, the first instance reaching the memory blocking phase may have to wait for a long period, this will result in a bigger downtime as the time for waiting, other memory pages could be updated easily. Also the CPU, in that case, can be exhausted as the actions are in a parallel fashion.

*2) Synchronized Round-Robin Pattern:* The Round-Robin strategy aims to reduce the migration CPU load, we achieve this through grouping by phase the steps of an SFC migration and then executing them in-order. This approach reduces consumption of system resources caused by an SFC migration, albeit at a significantly higher total migration time, as we do all actions sequentially and as previously, this SFC migration strategy uses all available network resources.

*D. Network-aware based Service Function Chain Migration*

5G networks are expected to support various URLLC's services, which requires strict delay constraint. However, none of the previous approaches can guarantee these prerequisites because of the randomized way for handling SFC migration when it comes to network resources. Indeed, having a huge number of applications capable of following and serving users can compromise all network resources among MECs by allowing a large number of migrations at the same time. We propose the network-aware based SFC migration to address these requirements, as its purpose is to refine the network usage, reducing the overhead, and enabling better users' QoE. By controlling the network's bandwidth, our network-aware based SFC migration triggers low-consumption migration operations across the networks. Initially, we gather all the available information on bandwidth and latency between each pair of MEC nodes, thus obtaining a global knowledge of the distributed infrastructure. Then, after each migration's decision is taken, given network resources are reserved to allow that migration and better usage of the global bandwidth. Finally, our network-aware solution frees the used resources as it completes SFC migrations. It is noticed that when using rsync, the whole bandwidth is used by one process. But upon other processes start a network transfer operation (either migration or simple traffic), the bandwidth will be shared among all the processes using the best-effort policy. Thus based on this observation, if we start "n" migrations simultaneously, the bandwidth usage will be shared among them. Through a qualitative assessment, we can observe that if "n" becomes too big, then the migration time will tend to the infinity, which will result in a fiasco to network operators. We also emphasize that we reserve bandwidth for each SFC migration based on the last iteration, that stops the container. Thus, if the reserved bandwidth offers a downtime transfer similar to when having the full utilization of the bandwidth then the reservation's limit is set.

*E. Implementation*

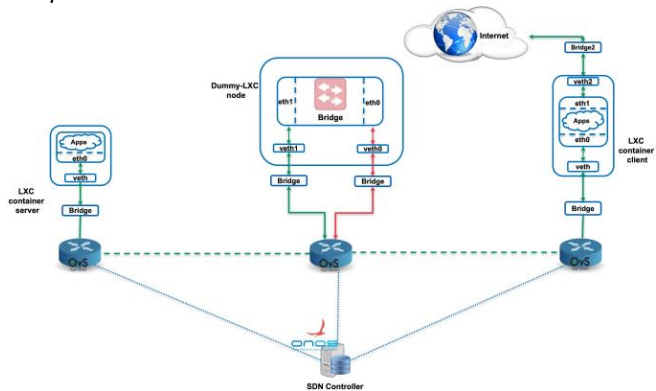

Fig. 2: Architecture of the implementation.

Enabling the migration of multiple independent virtual instances can be relatively easy when compared to migrating several instances with a close relationship (SFC). Initially, before enabling an SFC migration, the creation and the formation of an SFC is the starting point of interest. As a testing SFC implementation is unavailable, we design a service chain for a video streaming application. In this SFC, each video passes through an intermediate traffic redirection node, built on top of an LXC container and has two SDNenabled network interfaces. This redirection instance acts as a turnaround node in the network where the integrality of the traffic should go through in both directions, i.e., from the video server to the client and vice versa as depicted in Fig. 2. As stated before, our solution considers using the SDN paradigm thus, OVS switches are configured using the ONOS SDN controller, where each node in the SFC has its own switch. The only requirement is the availability of the virtualization software to deploy LXCs container engine and programmable switches OVSs. We use our previous proposed solution, presented in [13], to allow better isolation between incoming/out-coming traffic. To implement the turnaround logic in the Dummy-LXC host, a Bridge is created inside the Dummy-LXC container (the Red Bridge in Fig. 2). Two outer network interfaces, veth0 and veth1, respectively, are plugged in the outside OVS (the blue one in the same Figure); and inner network interfaces named eth0 and eth1 are attached to the red Bridge in their turn.

## IV. EXPERIMENTAL EVALUATION

We experimentally evaluated our proposed SFC migration schemes using one physical server. The server has 48 cores with VT-X support enabled, 256 GB of memory, 1Gbps interconnection, and Ubuntu 16.04 LTS with the 4.4.0-77-generic kernel and QEMU-KVM installed. Two virtualized computer nodes were used to evaluate the proposed implementation. Each one representing a different MEC node (i.e. a DN node in ETSI and 3GPP proposals). The first VM is acting as a source DN and the latter VM represents the target DN. Each DN uses Ubuntu 16.04 LTS with the 4.4.0-64-generic kernel and has 16 virtual core CPU and 32GB of main memory. The container environment was setup using LXC 2.8 and CRIU 3.11. It is noticed that two additional hosts were used for the management plane. The first host acts as an SDN controller that manages the communications between the different DNs. As SDN controller, we used ONOS, however, any other SDN controller could be used as well. While the second host serves as a global orchestrator for handling the life-cycle of SFCs (i.e. from the creation phase till the migration or the deletion stage). It is noteworthy that the global orchestrator uses an enhanced version of MIRA!, a framework previously presented in [13], that support our proposed SFC migration patterns and schemes.

We start by evaluating the Asynchronous SFC migration pattern under diverse network bandwidth limitations to select the most appropriate bandwidth limit for reducing SFC's migration overhead. In that evaluation, both the downtime and the total migration time will be analyzed and discussed. Finally, based on satisfactory bandwidth usage, a CPU consumption analysis will be presented to compare all the approaches introduced and detailed earlier in Section III.

For each SFC migration scheme, we conducted a set of experiments evaluating both the downtime and total time under various network configurations and CPUs' load; each was repeated ten times. The SFC evaluated is consisting of a video server streamer offering videos on demand (VoD) to clients passing through an intermediate node dubbed DummyLXC node that forms our second virtualization instance to be migrated when the SFC migration is triggered. The DummyLXC, and video server nodes sizes are equal to 470 MB and 573 MB respectively.

### A. Downtime Analysis

Fig. 3 depicts the induced downtime under diverse network configurations. The main purpose of this experiment is to optimally exploit network resources and avoid breaking down the whole network infrastructure. The detailed explanation on how this phenomenon can occur was introduced prior in subsection III-D. This experiment outputs the downtime, standard deviation, 95% confidence interval (CI), and coefficient of variation (CV) results for both elements constituting our developed service chain considering various bandwidth values as part of defining the most suitable network configuration. Detailed values are presented in Table I. It is noticed that we used the Asynchronous SFC migration pattern to compute a reasonable bandwidth limit as this pattern represents the empiric approach due to the absence of control in that SFC migration scheme. As expected, the results for the videostreaming container are larger when compared to the DummyLXC container results for all bandwidth values. The difference in these results is due to the additional copies of the network connections status.

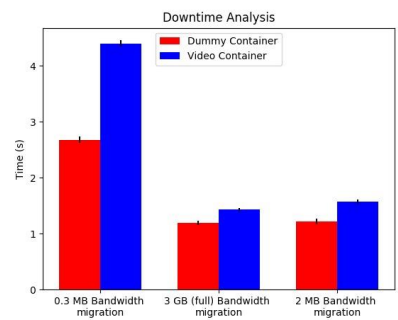

Fig. 3: Downtime comparison for the Asynchronous SFC Migration pattern under diverse network configurations.

Meanwhile, the full Bandwidth (i.e. 3 GBps) represents the maximum available bandwidth between two DNs (i.e. second error bar in Fig. 3). The maximal bandwidth value was set using the IPerf tool [17], measures were taken ten times, and the collected values were averaged to obtain the mean bandwidth. This case should deliver the best results in terms of downtime and total migration time when fully exploited by instances forming the migrated SFC. However, limiting the bandwidth to SFC migration processes in 5G networks will allow better exploitation of network resources in case of a massive number of migrations. Yet, choosing the right value is a challenging process, as a low bandwidth can increase both the total migration time and the downtime causing damage to the migration process. While an overestimated threshold will waste network resources in vain. In Fig. 3, we selected three bandwidth values for testing the downtime efficiency. The full bandwidth usage is taken as a reference and at the same time the overestimated value since it is the bigger value and the one offering best results in case of lack of overhead. While 0.3 MBps (i.e. first error bar in Fig. 3) is the underestimated value and 2 MBps value is the satisfactory value (i.e. error bar number three in Fig. 3). It should be pointed out that the 2 MBps value was obtained by trying many bandwidth values with one constraint in mind, which is having similar/near results to the full bandwidth utilization. Based on Fig. 3 and Table I, we can derive that the downtime for the Asynchronous SFC migration pattern is quite similar for both the full bandwidth and the 2 MBps migration bandwidth. The obtained value represents a reduction of 99.93% from the initially provided bandwidth without affecting downtime results. While if selecting the 0.3 MBps value, an increase of three times the full value will be observed.

TABLE I: Downtime comparison in case of different Bandwidth values.

| Bandwidth (Asynchronous SFC Mig.) | Mean Time (s) | Std deviation | CI 95% | Coef Var |
|---|---|---|---|---|
| Dummy-LXC 0.3 MB | 2.674 | 0.075 | 0.056 | 0.028 |
| Video server 0.3 MB | 4.397 | 0.076 | 0.057 | 0.017 |
| Dummy-LXC 3 GB | 1.189 | 0.049 | 0.037 | 0.041 |
| Video server 3 GB | 1.429 | 0.047 | 0.035 | 0.033 |
| Dummy-LXC 2 MB | 1.222 | 0.066 | 0.05 | 0.054 |
| Video server 2 MB | 1.571 | 0.056 | 0.042 | 0.036 |

*B. Total Time Analysis*

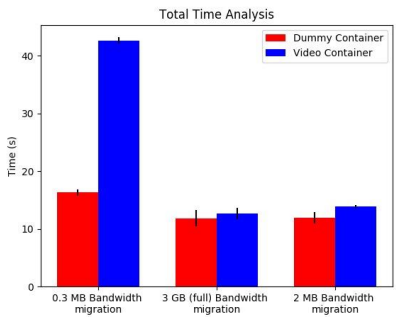

Fig. 4: Total migration time experienced for the Asynchronous SFC Migration pattern under diverse network configurations.

To strengthen our assumptions related to limiting network resources so that more efficient SFC migration schemes will be admitted, we extended our evaluation to cover total migration time under different bandwidth configurations. We addressed this evaluation using the same experimental scenarios of this section and plot the results in Fig. 4 for the Asynchronous SFC migration pattern (i.e. SFC is composed by the Dummy-LXC (red) and video-streaming (blue) containers). In Fig. 4, the Y-axis is in seconds and for each bar, we also plotted the 95% CI of the mean.

The mean total migration time, the Std deviation, the 95%CI, and the CV for SFCs under different network configurations are presented in Table II. As expected the full bandwidth (i.e. (3GB) second error bar in Fig. 4) and the 2 MBps (last error bar in Fig. 4) scenarios were quite similar, while the 0.3 MBps case increased approximately four times the total migration time than the expected value. It is important to note that to get a comparable value between the full bandwidth case and the 2 MBps case, we leveraged our work [18] that optimizes the disk copy otherwise the value of the bandwidth must be increased as more data need to be transferred over the network.

From the results, we can also observe that for all bandwidth configurations the video-streaming container takes longer than the Dummy-LXC one. This additional time is logical as initially, the video-server has a bigger size when compared to the Dummy-LXC. Furthermore, for the video-server container, the longer migration time in comparison with the DummyLXC one is due to the greater number of memory pages being copied. Thus, we can conclude that the overall total migration time of the SFC will increase as the two instances are dependent. However, other SFC migration patterns will be considered and investigated in terms of CPU load for a better approach in the next sub-section.

TABLE II: Total time comparison in case of different Bandwidth values.

| Bandwidth (Asynchronous SFC Mig.) | Mean Time (s) | Std deviation | CI 95% | Coef Var |
|---|---|---|---|---|
| Dummy-LXC 0.3 MB | 16.296 | 0.667 | 0.503 | 0.041 |
| Video server 0.3 MB | 42.658 | 0.742 | 0.56 | 0.017 |
| Dummy-LXC 3 GB | 11.833 | 1.891 | 1.426 | 0.16 |
| Video server 3 GB | 12.661 | 1.273 | 0.96 | 0.1 |
| Dummy-LXC 2 MB | 11.922 | 1.28 | 0.965 | 0.107 |
| Video server 2 MB | 13.842 | 0.346 | 0.26 | 0.025 |

*C. CPU Consumption Analysis*

The CPU consumption analysis experiment was conducted to allow a better understanding of all the proposed SFC migration patterns. Based on the two previous analysis, the bandwidth was set to 2 MBps. Fig. 5 illustrates the variation of CPU loads following three types of SFC migration schemes mainly Asynchronous, Synchronized (Wait-For-Me) and Synchronized (Round-Robin) SFCs migrations. In Fig. 5, the Y-axis represents the CPU's load percentage in the source DN node, and the X-axis portrays 100 seconds sample of time in seconds where the SFC migrations occur. For the Asynchronous SFC migration pattern, the red color is used to represent the

CPU variation during the migration process. Meanwhile, the grey and blue colors are chosen to express Synchronized (Wait-For-Me) and Synchronized (RoundRobin) SFCs respectively. In Fig. 5, before 20 seconds and posterior to 75 seconds periods outline before starting the SFC migration and after achieving the SFC migration respectively. Meanwhile, in the range between 21 and 74 seconds, we can observe that while the Asynchronous SFC migration pattern has the fastest migration time (i.e. from 35s to 45 seconds, this also can be verified leveraging the previous total time graph in Fig. 4 and the Table. II respectively), it induces the highest CPU load. The Synchronized (Wait-For-Me) SFC migration pattern is the symmetrical approach as it stresses the CPU considerably while not consuming a lot of time during the SFC migration, showed with grey color in Fig. 5. This pattern takes 2 to 3 supplementary seconds compared to the Asynchronous approach. This additional time is due to the increase in the downtime as a consequence for waiting for the second container to reach the last iteration phase. It should be mentioned that this time may increase more when augmenting SFC's components number, especially if they have different disk sizes. Finally, the Round-Robin Synchronized approach consumes the less CPU overhead among all other patterns. It is quite similar to a subsequent migration scheme except the Round-Robin policy is applied within the decomposed parts of the migration between various components of the SFC.

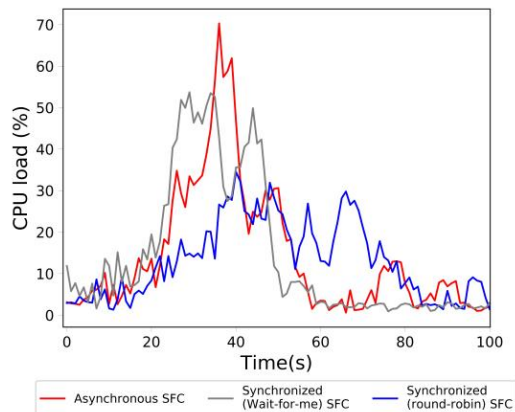

Fig. 5: CPU consumption analysis in case of different SFC migration patterns.

However, this approach takes longer compared to previous patterns in term of total migration time.

*D. Results Discussion*

Based on the observations gathered from the previous subsections related to the downtime, total migration time, and CPUs' loads we surmise that there is no clear winner in performance. Thus, the right SFC migration pattern must be selected based on users' motion, applications' requirements and MEC nodes' resources. Furthermore, the network-aware SFC migration pattern is selected to act as a support for the Synchronized (Wait-For-Me), Synchronized (Round-Robin) and the Asynchronous SFC migration patterns considering the delicacy of 5G networks in terms of the number of users and available resources (network or system resources). For instance, a combination of the Round-Robin approach and the network-aware SFC patterns is accepted when users' path is known and we can proactively plan their trajectories. This will reduce the CPU's overhead and optimize network resource wastage. While the Synchronized Wait-For-Me pattern could be exploited with the network-aware SFC pattern to handle applications that do not require ultra-low latency as the WaitFor-Me approach doesn't guarantee the lowest downtime.

V. CONCLUSION AND FUTURE WORK

In this work, we designed, proposed and evaluated four SFC migration patterns for allowing the support of synchronized depending applications (SFC or state-full micro-services based applications). The obtained results showed that there is no clear winner in our presented patterns, thus, a trade-off or hybrid combinations are the favorite proposals. Additionally, we have shown that the network-aware SFC pattern should act as a support for the other proposed patterns. Our future work will focus on employing Reinforcement Learning (RL) techniques to bypass the brute force search method. Along with extending the same RL agent to allow dynamic manageability of the network's bandwidth irregularity.

ACKNOWLEDGMENT

This research work is partially supported by the European Union's Horizon 2020 research and innovation program under the MATILDA project with grant agreement No. 761898. It is also partially funded by the Academy of Finland Projects CSN and 6Genesis under grant agreement No. 311654 and No. 318927, respectively.